\documentclass[12pt]{iopart}

\usepackage{graphicx} 
\usepackage{color}
\begin{document}

\title[The Er site in nanosized CaF$_2$]{The incorporation site of Er in nanosized CaF$_2$}

\author{F. d'Acapito}
\address{CNR-IOM-OGG, c/o ESRF, F-38043 Grenoble France}
\ead{dacapito@esrf.fr}
\author{S. Pelli-Cresi}
\address{University of Bologna, Phys. Dept., I-40127, Bologna Italy}
\author{W. Blanc, M. Benabdesselam, F. Mady}
\address{Nice Sophia Antipolis University, CNRS, LPMC UMR 7336, 06108 Nice Cedex 2, France}
\author{P. Gredin$^*$, M. Mortier}
\address{Chimie ParisTech, CNRS, Institut de Recherche de Chimie Paris, PSL Research University, Paris, France}
\address{$^*$ also at Sorbonne University, UPMC Univ. Paris 06, Paris, France}
\begin{abstract}
The incorporation site of Er dopants inserted at high and low concentration (respectively 5 and 0.5 mol \%) in nanoparticles of CaF$_2$ is studied by X-ray Absorption Spectroscopy (XAS) at the Er L$_{III}$ edge. The experimental data are compared with the results of structural modeling based on Density Functional Theory (DFT). DFT-based molecular dynamics is also used to simulate complete theoretical EXAFS spectra of the model structures. The results is that Er substitutes for Ca in the structure and in the low concentration case the dopant ions are isolated. At high concentration the rare earth ions cluster together binding Ca vacancies. 
\end{abstract}

\maketitle
\section{Introduction}
Rare Earth (RE) nanoparticles (NPs) have recently attracted a considerable interest due to their optical properties \cite{Mai2006}. In particular, NPs of Sodium-based fluorides of the class Na(RE)F$_4$ are reported as efficient systems for upconversion, particularly appreciated for bioimaging \cite{Bouzigues2011, Nyk2008, Kumar2007}. Among fluorides, CaF$_2$ constitutes a particularly interesting material for its stability and a non hygroscopicity. 
Dy - and Tm- doped CaF$_2$ showed very interesting dosimetric properties. Ionizing radiation dosimeters TLD200 and TLD300 were respectively  manufactured and marketed as thermoluminescent dosimeters (TLDs). 
CaF$_2$:Tm is particularly interesting for mixed (n, g) field dosimetry because of the different sensitivity of its thermoluminescent peaks to neutrons and to gamma rays \cite{Bos1991}.
Er-doped CaF$_2$ is reported as an efficient material for lasers at 2.79 $\mu$m \cite{Labbe-02}
 whereas Yb-doped CaF$_2$ is reported in applications for tunable laser \cite{Petit2004, Aballea-15, Akchurin-13} and Er-doped CaF$_2$ NPs are reported as promising materials for applications in the third window of low absorption for silica based optic fibers \cite{Bensalah2006}. \\
CaF$_2 $ cristallizes with the Fluorite structure \cite{Smakula1955} consisting in a cubic arrangement of F$^-$ ions with Ca$^{2+}$ occupying the center of alternate cubes. Er$^{3+}$ substitutes for Ca$^{2+}$ in the structure and, to balance the excess of positive charge, it is associated to a charge balancing defect. The structural properties of trivalent dopants in CaF$_2$ single crystals have been investigated in the past using Electron Paramagnetic Resonance \cite{McLaughlan1966}, Ionic Thermo Current \cite{Stott1971, Kitts1974}, Audio-Frequency Dielectric Relaxation \cite{Fontanella1976, Andeen1979}.  In the case of low-doped ($\leq 0.1 Mol \%$) samples it has been established that the principal complex consists in (adopting the Kroger-Vink notation) a Er$_{Ca}^{\bullet}$+F$_I^{'}$ where F is placed in a nearest (called nearest neighbor configuration, NN) free cube center respect to Ca. Minor contributions were proposed to be due to bonds to oxygen ions and complexes with the F ion in a next-nearest neighbor (NNN) positions \cite{Edgar-75}. For higher concentration values (a few \%) theoretical calculations \cite{Bendall1984, Corish1982} foresee complexes involving aggregates of REs substituting for Ca associated to F interstitials. Direct structural investigations on the RE site based on X-ray Absorption Spectroscopy (XAS) \cite{Catlow1984} permitted to confirm the theoretical calculations in particular by observing in the case of Er: 9 F ions at 2.35 \AA \thinspace and 8 Ca ions at 3.93 \AA. \\
In the present study XAS has been used to determine the incorporation site of Er in CaF$_2$ nanoparticles, i.e. in a system not produced with the high equilibrium procedures as the single crystals. 
\section{Experimental}
\subsection{Sample preparation}
All syntheses were made using commercially available nitrates: $Ca(NO_{3})_{2},4H_{2}O$ 99.98\% and  $Er(NO_{3})_{3},5H_{2}O$ 99.99 \% from Alfa Aesar. 48 wt\% hydrofluoric acid (Normapur Prolabo) is used as the fluorinating agent. Particles of  erbium doped calcium fluoride are obtained using a co-precipitation method. A solution containing the cationic precursors in stoichiometric proportion, was made by dissolving nitrate salts in deionized water. Then this solution was added dropwise to the hydrofluoric acid solution which was stirred magnetically, leading to the formation of $Er^{3+}$ doped $CaF_{2}$ particles according to the following theoretical reaction:
\\
$(1-x)Ca(NO_{3})_{2} + xEr(NO_{3})_{3} + (2+x)HF  \longrightarrow Ca_{1-x}Er_{x}F_{2+x}$
\\
The hydrofluoric solution contains a large excess of $F^{-}$ anions (10 times the stoichiometric quantity needed) in order to keep a concentration of fluorine anions approximately constant during the reaction. \\
The obtained mixture was centrifugated at 13,000 rpm for 20 minutes. The recovered particles were then washed and centrifugated with deionized water 4 times before drying  at 80$^{\circ}$ C. Two batches of powder were prepared with Er content at 0.5 mol \% and 5 mol \%. The obtained powder was annealed at 350$^{\circ}$ C under argon atmosphere for 3h. 
\subsection{X-ray Diffraction} 
\begin{figure}
\includegraphics[width=0.8\textwidth]{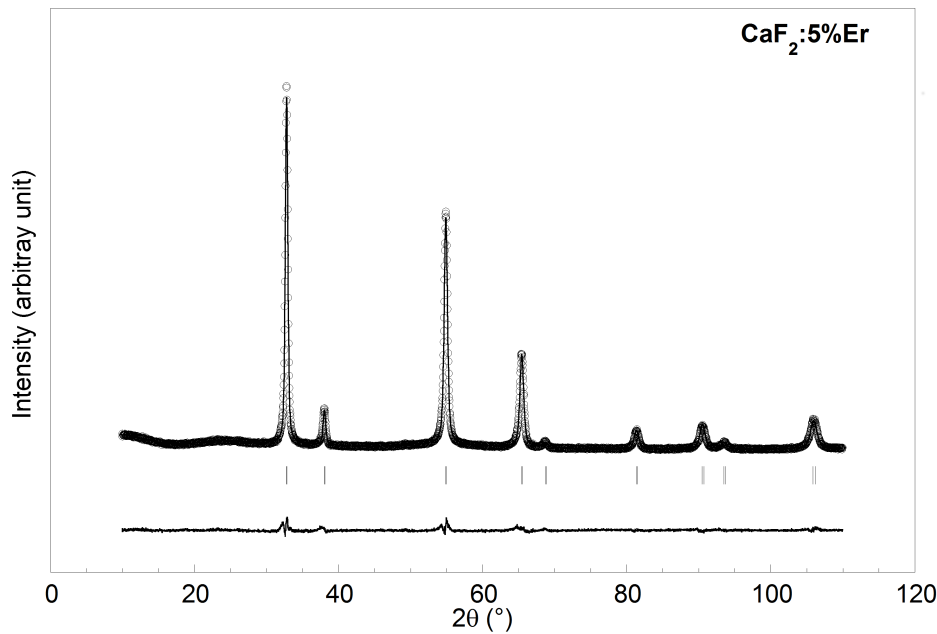}
\caption{\label{fig:xrd} XRD pattern refinement of $CaF_{2}:5\%Er$. Circles correspond to the observed profile and full curve to the calculated one. The short vertical lines below the profile curves mark the positions of all Bragg reflections. The lower curves show the difference between observed and calculated profiles.}
\end{figure}

The samples were analyzed using powder X-ray diffraction (XRD). Measurements were performed on a Bruker D8 Endeavour using a Co-radiation diffraction ( $\lambda_{K\alpha1} = 1.78892$ \AA \thinspace  and $\lambda_{K\alpha2} = 1.79278$ \AA \thinspace ). The 2$\theta$ angular resolution was 0.029$^{\circ}$. The diffraction patterns were scanned over the 2$\theta$ range 10 - 110$^{\circ}$ (Fig.\ref{fig:xrd}).
The XRD patterns of the particles present peaks which can be indexed in $CaF_{2}$ cubic phase of the fluorite-type structure (space group Fm3m). XRD patterns present broad peaks characteristic of small crystallite size with a mean value $\Lambda$ calculated using the facilities of the refinement program Fullprof \cite{Rodriguez-01}.
%
After correction for the instrumental broadening the width of the peaks is similar in the two cases (5 \% and 0.5 \% doped) evidencing that this parameter is dominated by the reduced crystallite size. This fact prevents the detection of possible fine lattice distortions caused by the dopants and associated defects especially in the highly doped sample. The lattice parameter measured was a$_{0.5\%Er}$=5.471(2) \AA \thinspace and a$_{5\%Er}$=5.493(2)  \AA \thinspace corresponding to an expansion of respectively 0.1\% and 0.5\% respect to the literature value (a=5.46342(2) \AA \cite{smakula-55}). The size of the crystallites derived from the peak width values was found to be $\Lambda \approx$ 30 nm and this value has been confirmed by Transmission Electron Microscopy which evidenced an average particle size of 25nm.
\subsection{X-ray Absorption spectroscopy measurements} 
X-ray Absorption Spectroscopy (XAS) \cite{RevModPhys.53.769} measurements at the Er-L$_{III}$ edge were carried out at the LISA (formerly GILDA) beamline at the European Synchrotron Radiation Facility in Grenoble \cite{dacapito-14}. The monochromator was equipped with a pair of Si(311) crystals and was run in Dynamically focusing mode. Pd-coated mirrors (E$_{cutoff}$=19 keV) were used for vertical collimation, focusing and harmonic suppression. XAS data  were collected at Liquid Nitrogen Temperature in Fluorescence detection mode using an hyper-pure Ge 12-elements detector array for the most diluted samples and in transmission mode for the others. Samples were obtained by finely grounding the powder, mixing to cellulose binder and pressing in pellets. The pellets were sufficiently transparent to allow the collection of the absorption spectrum of a $Er_2O_3$ reference after them allowing a precise control of the energy calibration during the data collection. For each sample 2 to 4 spectra were collected and averaged in order to improve the Signal to Noise ratio. 
\section{Ab-Initio Structural Modeling and Molecular Dynamics}
Theoretical methods were used to simulate the structure around the Er ions, to estimate the formation energy of different defects and to simulate full XAS spectra via Molecular Dynamics (MD). 
%
%
The structure of different point defects with Er in CaF$_2$ were calculated with Density Functional Theory (DFT) as implemented in the VASP code \cite{kresse-96} on rhombohedral supercells based on the $CaF_2$ structure. The choice of the dimension of the supercell is a tradeoff between the need of obtaining a sufficiently large cell to mimic an isolated defect ($2Er_{Ca}^{\bullet}$+$V_{Ca}^{"}$ has a dimension of about 6.5 \AA) and the need of keeping the computational time at a reasonable value. A $2\times2\times2$ cell would have been too small (side 7.7 \AA \thinspace barely separating the defects) whereas the $5\times5\times5$ would have contained an untreatable number of atoms (375). For the DFT calculations (involving a high number of points in the mesh of the reciprocal $k$ space) $3\times3\times3$ rhombohedral supercells containing about 81 atoms (side about 11.6 \AA) were used. For the DFT-MD calculations, using just the $\Gamma$ point in $k$ space but needing large clusters to calculate EXAFS signals at large distances, $4\times4\times4$ cells of size 15.5 \AA \thinspace and a total 192 atoms were used.
Calculations were done with projector augmented wave (PAW) pseudopotentials and the exchange-correlation functional used was the generalized gradient approximation (GGA) \cite{96-prl-perdew}. Plane waves were considered with a cut-off energy of 650 eV. The Brillouin zone was sampled using a $4\times4\times4$ k-point mesh (using the Monkhorst-Pack scheme \cite{76-prb-monkhorst}). At each ionic step, the electronic structure was optimized until attaining a convergence of the total energy within $10^{-6}$ eV, whereas the ionic positions were optimized until Hellman Feynman forces were below $10^{-4}$ eV/\AA.The validity of the procedure was tested on simulating some well known compounds and comparing the obtained lattice parameters with those available from literature. For the simulated CaF$_2$ structure a lattice parameter a$_{CaF2}^{DFT}$=5.4982 \AA \thinspace was found against a bibliography \cite{smakula-55} value of a$_{CaF2}^{Bib}$=5.4634 \AA \thinspace meaning that in our case the value found was only 6/1000 expanded than the experimental value. The same test carried out on the orthorhombic phase of ErF$_3$ \cite{kraemer-96} yielded changes of 0\%, -3\%, +7\% of the a, b and c axes lengths respect to the literature values.
%
The point defects considered here were a "substitutional" $Er_{Ca}^\bullet$, a charge compensated "double substitutional + Ca vacancy" $2Er_{Ca}^{\bullet}$+$V_{Ca}^{"}$ and the "substitutional plus F interstitial" $Er_{Ca}^\bullet$+$F_{I} ^{'}$, following the standard Kroger-Vink notation. In the latter cases the charge compensating vacancy/interstitial was either associated to the Er ions (placed as near as possible: F nearest anionic neighbor, Ca first cationic neighbor) or dissociated (placed as far as possible in the supercell).
The details of the relaxed structures are shown in the last lines of Tab.~\ref{tab:exares}. The Er concentration levels were around 4-8 cationic \%; concerning the lattice parameters these simulations confirm the lattice expansion found experimentally although at a lower level (0.06\%-0.2\% depending on the particular complex).  \\
In order to state the possibility of finding these complexes in the crystal the associated formation energies were calculated. The elemental chemical potentials $\mu_{Ca}$, $\mu_{Er}$, $\mu_{F}$ were derived from the formation energies of compounds  ErF$_3$, Ca$F_2$ considered in equilibrium with F$_2$ gas, following a procedure already adopted in previous works \cite{DAcapito2013}. The formation energies of the complexes are calculated as the difference between the sum of the chemical potentials and the total energy provided by VASP. A collection of results is presented in Tab.~\ref{tab:formEn}. 
\begin{table*}[ht]
\caption{\label{tab:formEn} Formation energies of the different complexes considered. The errors are estimated from the variation of the final energy value for different k meshes and cutoff energy values and fixed to 0.01 eV. }
\small
\begin{tabular}{ll}
\hline
Complex  &  Formation energy (eV)  \\
\hline
Er$_{Ca}^{\bullet}$                                          &  7.04 \\
2Er$_{Ca}^{\bullet}$+V$_{Ca}^"$ assoc.          &  0.88\\
2Er$_{Ca}^{\bullet}$+V$_{Ca}^"$  dissoc.        &  1.21 \\
Er$_{Ca}^{\bullet}$+F$_{I}^{'}$ assoc.            &  1.04 \\
Er$_{Ca}^{\bullet}$+F$_{I}^{'}$   dissoc.         &  1.09 \\
\hline
\end{tabular}
\end{table*}
In the case of charge-compensated sites the structure around Er presents several bond distance values that make difficult a direct comparison with the results of the EXAFS analysis. The theoretical EXAFS signal generated by Feff can not be used as it does not calculate the Debye-Waller factors (DWf) that heavily influence the overall shape of the signal. In order to obtain a realistic comparison with experimental data a DWf must be taken into account and for this reason a simulation of the EXAFS spectra via Molecular Dynamics (MD) was carried out. Following the procedure presented in \cite{palmer-96} MD trajectories for each defect were calculated using MD-DFT as implemented in VASP. The starting point was the relaxed structure obtained by 'static' DFT and then the MD was carried out in the canonical ensemble (NVT) with the temperature stabilized at 300 K via a Nose thermostat. 
%
Atomic steps were carried out every 1fs  the total relaxation time was about 300 fs. The last 150 frames were used to calculate the XAS signals with the Feff9.6.4 code \cite{Rehr-10} using theoretical signals, calculated via a Self-Consistent procedure,  and averaged. The residuals between N averaged spectra $\Theta_N $ and N-1 (defined as $\xi = \sqrt {\sum_k (\Theta_N^2 - \Theta_{N-1}^2)} $ were a few $10^{-3}$ units. The obtained spectra were compared to those obtained from a smaller cell ($3\times3\times3$, 81 atoms) and longer evolution times (1ps) finding an identical damping of the main signals, meaning that the present averaging time was sufficient to correctly reproduce the spectrum.
The comparison of the experimental data with the simulated spectra can be extremely effective in the determination of a dopant site in a matrix as shown in previous works \cite{cartechini-11}. 
\section{Results}
\begin{figure}
\includegraphics[width=0.5\textwidth, angle =-90]{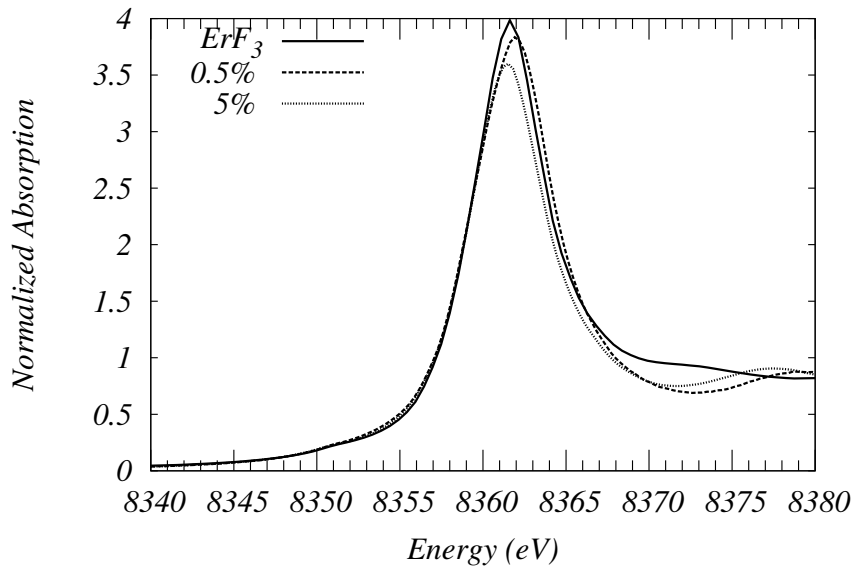}
\caption{\label{fig:allxan} Comparison of the XANES spectra of the lightly doped (dashed line) and heavily doped (thin dashed line) samples compared with the spectrum of ErF$_3$ (full line).}
\end{figure}

\begin{figure}
\includegraphics[width=0.5\textwidth, angle =-90]{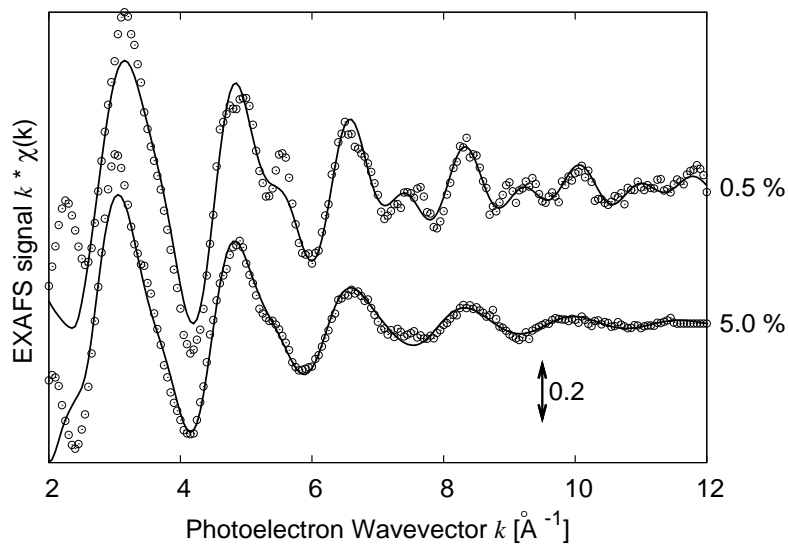}
\caption{\label{fig:allexa} Comparison of the EXAFS spectra of the samples (dots) and best fits.The spectra are vertically shifted for clarity and the vertical scale is shown by the double-arrow marker.}
\end{figure}

\begin{figure}
\includegraphics[width=0.5\textwidth, angle =-90]{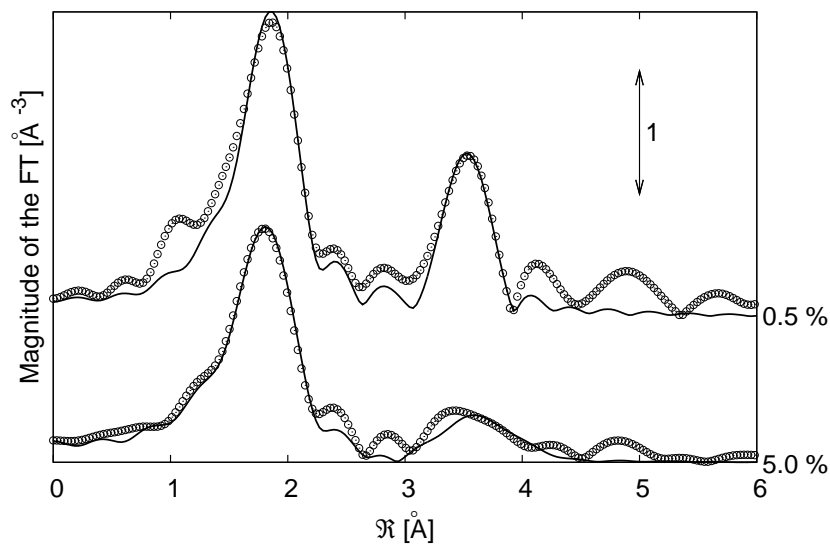}
\caption{\label{fig:allfou} Comparison of the Fourier Transforms of the EXAFS spectra of the samples (dots) and best fits. Both transforms were carried out in the interval k=[2.5-10.5]\AA$^{-1}$ with a $k^2$ weight and a Hanning window with apodization $\Delta k=0.1$  \AA$^{-1}$.The spectra are vertically shifted for clarity and the vertical scale is shown by the double-arrow marker.}
\end{figure}

The XANES spectra are shown in Fig.\ref{fig:allxan}
The edges of the samples and that of ErF$_3$ coincide so it can be derived that the state of Er ion is 3+. 
The EXAFS spectra of the different samples are shown in Fig.\ref{fig:allexa} whereas the related Fourier Transforms (FT) are shown in Fig.\ref{fig:allfou}.
Also in this case it can be noticed that the low-concentration Er-doped samples appear different from the highly concentrated ones. In particular high frequency components are visible in the related EXAFS spectra (sharp oscillations at $k \approx 5.5$ , $ \approx 7.5$, $ \approx 9.2$, $ \approx 10.1$ \AA$^{-1}$) that are strongly attenuated in the spectra of the high concentration samples. Correspondingly, in the FT a second shell peak is well visible at $R \approx 3.5$ \AA \thinspace in the spectra of the diluted samples that is strongly damped in the high concentration ones. The XAS data analysis was carried out by calculating theoretical XAS signals on a cluster (radius 6.75 \AA \thinspace, 98 atoms) derived from the structural model found by DFT for the Er$_{Ca}^\bullet$ point defect. Calculations were carried out using the Feff9.6.4 code \cite{Rehr-10} using a Hedin-Lunqvist approximation for the exchange potential. Scattering potentials were calculated via the Self Consistent procedure implemented in Feff using a cluster of 5.6 \AA \thinspace (50 atoms). Data extraction and modeling were carried out with the ATHENA/ARTEMIS codes \cite{Ravel:ph5155}.
Fitting was carried out in R space in the interval R=[1-4.4]\AA. The structure was reproduced with a two shell model Er-F and Er-Ca with associated shell radii R and Debye-Waller factors $\sigma^2$. Coordination numbers were fixed to the expected crystallographic values; the S$_0^2$ global amplitude parameter was fixed to 0.95 after a fit of an $Er_2O_3$ model compound. This structural model revealed to reproduce all the main features in k and R space of the spectrum (taking onto account noise on the data) with the exception of the sharp peak at $k \approx 5.5$ \AA \thinspace. 
%
Carrying out simulations of the XAS spectrum of the $Er_{Ca}$ complex from the static structures obtained by DFT it was possible to evidence that this peak is originated by the constructive interference of several signals at distances up to 9 \AA. As this value exceeds the maximum cluster dimension for the subsequent MD calculations, the fit of the structure at such distance was not attempted.
The R factor value is about 0.02 so indicating the goodness of the fit \cite{ifeffit-faq}.
The results of the quantitative EXAFS data analysis are shown in Tab.\ref{tab:exares}. 
\begin{table*}[ht]
\caption{\label{tab:exares} 
Upper panel : quantitative results of the EXAFS data analysis compared with the unperturbed $CaF_2$ crystal. Errors on the last digits are indicated in parentheses. Lower panel: Er-F and Er-Ca bond distances obtained from the DFT calculations on the various clusters around Er. Values differing for less than $\pm 0.02$ \AA \thinspace have been grouped and asociated to an average value.
 }
\small
\begin{tabular}{lllllll}
\hline
Sample  &  \multicolumn{3}{c}{F shell}                 & \multicolumn{3}{c}{Ca Shell} \\
\hline
             & N   & R (\AA) & $\sigma^2$(\AA$^2$) & N  & R (\AA)  & $\sigma^2$(\AA$^2$) \\
\hline
CaF$_2$ & 8 & 2.381      & -                                 & 12 & 3.888     &- \\
0.5 \%         & 8 & 2.27(2)    & 0.005(1)                      & 12 & 3.91(2)  & 0.003 (2) \\
5.0 \%        & 8 & 2.26(2)    & 0.007(2)                      & 12 & 3.93(4)  & 0.02 (1) \\
\hline
Er$_{Ca}^{\bullet}$                           & 8 & 2.282    & -                   & 12 & 3.92  & - \\
2Er$_{Ca}^{\bullet}$+V$_{Ca}^{"}$  & 2 &  2.19     & -                   & 4  & 3.87  & - \\
                                                         & 6 & 2.31      & -                   & 7  & 3.93  & - \\
Er$_{Ca}^{\bullet}$+F$_{I}^{'}$       & 4 & 2.28     & -                    & 4 & 3.72  & - \\
                                                         & 5 & 2.38    & -                     & 4 & 3.94  & - \\
                                                         &    &            & -                    & 4 & 4.02  & - \\
\hline
\end{tabular}
\end{table*}

\section {Discussion}
The XAS data provide a complete information on the chemistry and local structure around Er ions in nanostructured CaF$_2$. XANES reveals that Er is predominantly in the 3+ state as the absorption edge is in the same position as for ErF$_3$. The high and low concentration spectra present a clear difference in the high energy side of the white line and this aspect will be discussed later. The EXAFS quantitative analysis reveals that Er shrinks the structure at the local scale as the Er-F bond is considerably shorter than the corresponding Ca-F bond. The second shell on the other hand, remains at the same distance, meaning that the distortion is only limited to the first shell. No formation of oxides is evident as instead it was evidenced in previous studies on fluoride nanoparticles \cite{Fortes2014}. Comparing the results of the EXAFS analysis with the DFT static simulations it is evident that the $Er_{Ca}^\bullet$ and 2Er$_{Ca}^{\bullet}$+V$_{Ca}^{"}$ exhibit structural parameters ($<R_{Er-F}>$=2.28 \AA,$<R_{Er-Ca}>$=3.91 \AA ) well in agreement with the experiment. On the contrary the Er$_{Ca}^{\bullet}$+F$_{I}^{'}$ is not compatible with the XAS data as it presents an average value for the first shell Er-F distance $<R_{Er-F}>$=2.33 \AA \thinspace that is appreciably longer than the experimental value. It is worth noticing that this value is extremely close to what observed by XAS and calculated in Ref.\cite{Catlow1984} ($R_{Er-F}$=2.35 \AA) demonstrating the consistency of the present ab-initio calculations with those presented there and that in our case the incorporation site of Er is different from what observed in bulk crystals. This fact is moreover confirmed by the comparison of the present XAS spectra with those of Ref.\cite{Catlow1984} where the peak at 5.5 \AA$ ^{-1} $ is a faint shoulder. Isolated substitutional or coupled with Ca vacancies are then the best candidates for the incorporation site for Er at low concentration. \\
These data are supported by the analysis of the formation energies of the various complexes. If the charged $Er_{Ca}^\bullet$ complex possess a considerable formation energy ($\approx$ 7 eV) the presence of a V$_{Ca}^"$ lowers considerably this value to 1.21 eV. This means that even in the case of \textit{isolated} Er$_{Ca}^{\bullet}$ defects they must be associated (though not closely bound) to charge compensating Ca vacancies. The close association of these two centers leads to a further lowering of the energy (0.88 eV)  that results to be even slightly lower than that of the Er$_{Ca}^{\bullet}$+F$_{I}^{'}$ complex (1.04 eV in the associated configuration). This confirms that complexes involving Ca vacancies can be obtained in the CaF$_2$ crystal and that their formation energy is about the same value of that involving a F interstitial. \\
A clearer vision about the structure of the complexes can be obtained from the XAS spectra simulated by Molecular Dynamics. Indeed, binding a charge compensating defect leads to a considerable disordering of the local environment (see Tab.\ref{tab:exares}) with the appearance of subshells that in principle should lead to a reduction of the related EXAFS signal.
\begin{figure}
\includegraphics[width=0.5\textwidth, angle =-90]{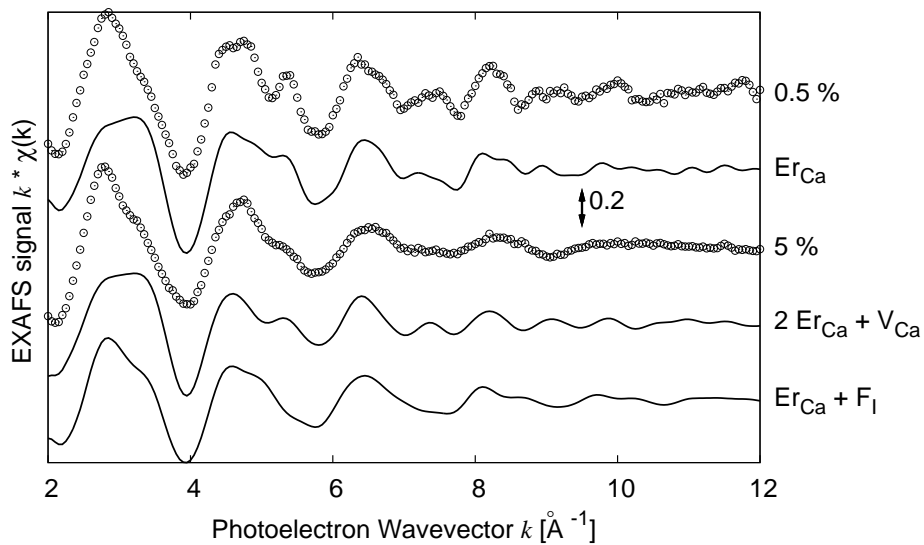}
\caption{\label{fig:allexa_mdComp} Comparison of the EXAFS spectra of the samples (dots) with the simulations of the spectra obtained by Molecular Dynamics for different sites (lines). %
A correction of 5 eV (close to the value of 4 found in the fits) was applied to the $E_0$ value of the experimental spectra to match the theoretical ones.
}
\end{figure}
%
%

\begin{figure}
\includegraphics[width=0.5\textwidth, angle =-90]{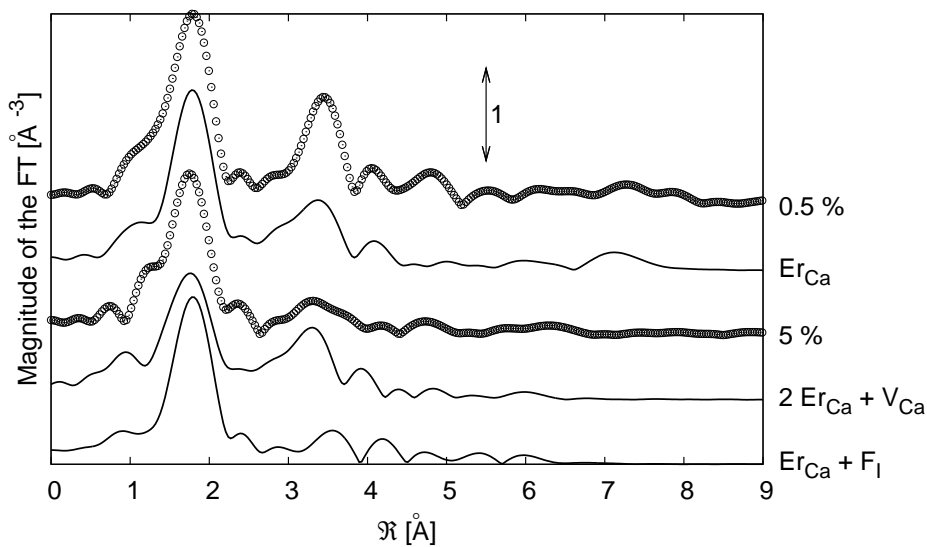}
\caption{
%
\label{fig:allfou_mdComp} Comparison of the Fourier Transforms of the EXAFS spectra of the samples (dots) with the simulations of the spectra obtained by Molecular Dynamics for different sites (lines). As in Fig.~\ref{fig:allfou}, the transforms were carried out in the interval k=[2.5-10.5]\AA$^{-1}$ with a $k^2$ weight and a Hanning window with apodization $\Delta k=0.1$  \AA$^{-1}$.
}
\end{figure}
\begin{figure}
\includegraphics[width=0.5\textwidth, angle =-90]{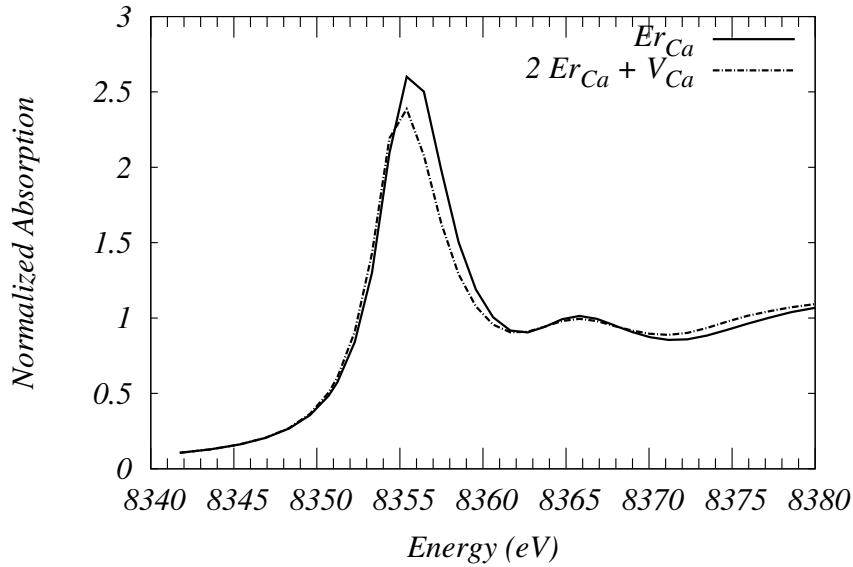}
\caption{\label{fig:allxan_dft} Comparison of the XANES spectra of the Er$_{Ca}^{\bullet}$ and the 2Er$_{Ca}^{\bullet}$+V$_{Ca}^{"}$ complexes. Spectra were calculated with the Hedin-Lunqvist exchange-correlation potential in the Full Multiple Scattering formalism on clusters of 98 atoms i.e. about 7 \AA \thinspace around the Er absorber.}
\end{figure}

This is exactly what observed for the second coordination shell in the samples with high Er content. \\
The comparison with the MD-DFT ( Fig.\ref{fig:allexa_mdComp} and Fig.\ref{fig:allfou_mdComp}) simulated spectra provides a deeper insight in this issue. Indeed, the simulated EXAFS spectra for Er$_{Ca}^{\bullet}$ and 2Er$_{Ca}$+V$_{Ca}$ present some features ( shoulders of the main oscillation at $k \approx 5.5$ \AA$^{-1}$ and $k \approx 7.5$ \AA$^{-1}$) that are absent in the simulated spectrum of 2Er$_{Ca}$+F$_{I}$. We could then privilege the first two as the possible incorporation sites for Er, and this in agreement with the conclusion drawn from the analysis of the static DFT data. The cited shoulders are more evident in the spectrum of Er$_{Ca}^{\bullet}$ and appear dampened in the spectrum of 2Er$_{Ca}$+V$_{Ca}$, reproducing qualitatively the situation observed upon increase of Er concentration in the crystal. It can be then stated that the increase of Er content leads to a clustering of dopants with charge compensating defects forming neutral complexes with appreciable disorder namely in the second coordination shell.The possible presence of REs ions even in the second shell cannot be excluded (clustering of dopants has been already foreseen and observed, namely in \cite{satta-prb-05, seo-jpcm-13}) but could not be assessed due to the faintness of this signal in the high concentration sample. \\
The aggregation of REs is particularly interesting for applications in mid-infrared lasers (2.7 $\mu$m) involving the transition $ ^4I_{11/2} \rightarrow ^4I_{13/2} $. Indeed, the level lifetime reduction due to concentration is reported \cite{Labbe-02, Ma-16} to be greater for the $^4I_{13/2}$ respect to the $^4I_{11/2}$ meaning that the RE clustering facilitates the population inversion between these two levels. Indeed 4\% Er doped CaF$_2$ is reported \cite{Ma-16} to possess an optimal laser performance.
A further confirmation of the RE clustering can be derived from the simulation of the XANES spectra for the Er$_{Ca}^{\bullet}$ and the 2Er$_{Ca}^{\bullet}$+V$_{Ca}^{"}$c complexes (the calculation was carried out on the static structures) as shown in Fig.~\ref{fig:allxan_dft}.
The increase of intensity on the high energy side of the white line observed in the experiment (Fig.\ref{fig:allxan}) is reproduced here, confirming the idea that at high concentration the REs tend to cluster and that, from the considerations on the first shell distance and formation energy, they bind a charge compensating Ca vacancy. It cannot be excluded that the size of the aggregates could be even larger with formation of multimer RE structures as reported in Ref.\cite{Bendall1984}. However, considered the large size of the supercell that would be needed for the full simulation of the EXAFS spectra and the fact that EXAFS cannot see correlations beyond the Er-Ca shell, simulations on multimers were not carried out. The dimer presented here could be taken as a prototype of the other multimers.\\
This result presents a new scenario compared with the previous observations reported in literature where the role of F interstitials was shown to be predominant. Anyways, it must be taken into account that the nanometric size of the particles and the peculiar growth technique could play a role in the formation of defects in the crystalline matrix.

\section{Conclusion}
In this study the incorporation site of Er in CaF$_2$ crystals has been studied with a combination of experimental (X-ray Absorption Spectroscopy) and theoretical (density functional theory, DFT and molecular dynamics MD-DFT) methods in the case of high (5 \%) and low (0.5\%) concentration values for the dopant. Er is present in the matrix as a 3+ ion and its local structure depends on the concentration. In all cases Er substitutes for Ca in the matrix with the difference that at low concentration Er ions are isolated (i.e. far from other dopants and/or charge compensating defects) whereas at high concentration they cluster together binding a charge compensating Ca vacancy.  

\clearpage
\newpage

\section*{References}
\bibliographystyle{iopart-num}
\bibliography{ercaf2_r03s}
\end{document}